\documentclass[11pt]{article}

\usepackage[utf8]{inputenc}
\usepackage[T1]{fontenc}
\usepackage{lmodern}
\usepackage[american]{babel}
\usepackage[margin=1in]{geometry}

\usepackage{amsmath}
\usepackage{amsthm}
\usepackage{amssymb}
\usepackage{amsfonts}
\usepackage{mathtools}

\usepackage{algorithm}
\usepackage{algorithmic}

\usepackage{graphicx}
\usepackage{subcaption}
\usepackage{float}
\usepackage{booktabs}
\usepackage{tabularx}
\usepackage{multirow}
\usepackage[font=small]{caption}
\usepackage{enumitem}
\setlist{nosep}
\usepackage{authblk}

\newtheorem{proposition}{Proposition}

\usepackage{hyperref}
\hypersetup{
    colorlinks=true,
    linkcolor=blue,
    citecolor=blue,
    filecolor=magenta,
    urlcolor=blue,
    pdftitle={Pricing and Calibration of Bitcoin Inverse Options via the Rough Bergomi Model},
    pdfauthor={Riccardo Caruso}
}

\title{\textbf{Pricing and Calibration of Bitcoin Inverse Options\\ via the Rough Bergomi Model}}
\author[1]{Riccardo Caruso}
\affil[1]{Department of Mathematics ``Tullio Levi-Civita'', University of Padova.
\texttt{riccardo.caruso@studenti.unipd.it}}
\date{}

\begin{document}
\maketitle

\begin{abstract}
Bitcoin inverse options, traded on the Deribit exchange and settled in the underlying cryptocurrency rather than in fiat currency, combine extreme and genuinely rough volatility dynamics with a non-linear, currency-dependent payoff structure. This paper develops and empirically validates a pricing and calibration framework for these instruments based on the rough Bergomi (rBergomi) model of Bayer, Friz and Gatheral (2016). We adapt the rBergomi dynamics to the inverse payoff $\max(S_T-K,0)/S_T$, and implement and compare three computational pipelines that differ in the simulation scheme for the driving fractional Brownian motion (coarse-grid Cholesky vs.\ the Hybrid Scheme of Bennedsen et al., 2017) and in the Monte Carlo pricing estimator (plain log-Euler vs.\ the Mixed Estimator of McCrickerd and Pakkanen, 2018). The model is calibrated to thirty implied volatility surfaces extracted from Deribit trade data between May 2022 and March 2025, spanning seven major market-stress events and nine baseline regimes stratified by volatility level. The Hybrid+Mixed pipeline is simultaneously the most accurate (mean unweighted RMSE 22.83 percentage points, versus 41.76\,pp for the Cholesky+Euler benchmark) and the fastest (17 seconds per snapshot, a 20-fold speed-up). The calibrated Hurst exponent is consistently close to the lower bound of the search space ($H\approx0.01$--$0.06$ in most regimes), confirming that Bitcoin's volatility is genuinely rough, and calibration error scales approximately linearly with the level of at-the-money implied volatility (Pearson $r=0.89$).
\end{abstract}

\section{Introduction}
\label{sec:intro}

The cryptocurrency derivatives market has grown into one of the largest options markets in the world, with the Deribit exchange consistently commanding 85--95\% of global Bitcoin options volume and open interest regularly exceeding \$20 billion \cite{elad2025}. Every option traded on Deribit is an \emph{inverse option}: margin, premium and settlement are all denominated in Bitcoin rather than in US dollars. This convention, adopted for operational reasons --- it allows the exchange to run entirely outside the traditional banking system --- has a direct mathematical consequence. A standard call payoff $\max(S_T-K,0)$, when converted into Bitcoin at the prevailing terminal exchange rate, becomes $\max(S_T-K,0)/S_T$, a convex, currency-dependent transformation that has no analogue in conventional equity or FX options markets \cite{alexander2023,lucic2024}.

A second, independent feature of the Bitcoin options market is the extreme roughness of the underlying volatility process. Multifractal analyses of Bitcoin realised volatility consistently estimate a Hurst exponent $H\in[0.05,0.15]$ \cite{takaishi2020,takaishi2021,takaishi2025}, far below the value $H=0.5$ implied by any diffusive stochastic volatility model. Through the asymptotic relation of Bayer, Friz and Gatheral \cite{bayer2016}, $H<0.5$ produces an at-the-money implied volatility skew that explodes as $T^{H-1/2}$ when $T\to0$ --- a power-law divergence that the Heston model \cite{heston1993}, whose short-maturity skew converges to a finite limit, structurally cannot reproduce \cite{forde2009}.

This paper combines these two strands --- rough volatility and inverse payoff mechanics --- into a single calibration framework for Bitcoin inverse options, and provides what is, to our knowledge, the first systematic empirical comparison of competing rBergomi simulation/pricing pipelines on cryptocurrency option data. Our contributions are threefold. First, we specify the rBergomi model with a piecewise-constant forward variance curve estimated directly from the delta-neutral at-the-money term structure, reducing the calibration to the three shape parameters $(H,\eta,\rho)$. Second, we implement and benchmark three computational approaches --- \emph{Cholesky+Euler}, \emph{Hybrid+Euler}, and \emph{Hybrid+Mixed} --- that combine two simulation schemes for the driving Volterra process with two Monte Carlo pricing estimators. Third, we calibrate all three approaches to thirty implied volatility surfaces spanning May 2022 to March 2025, including the immediate neighbourhood of seven major market events (the LUNA/UST collapse, the FTX bankruptcy, the SVB banking crisis, the spot Bitcoin ETF approval, the fourth halving, the BTC \$100k milestone, and the Trump inauguration) plus nine baseline dates stratified by implied volatility regime.

The remainder of the paper is organised as follows. Section~\ref{sec:inverse} reviews the payoff and risk-neutral pricing structure of inverse options. Section~\ref{sec:rbergomi} introduces rough volatility and specifies the rBergomi model. Section~\ref{sec:simulation} develops the Hybrid Scheme and the Mixed Estimator and defines the three computational approaches. Section~\ref{sec:calibration} describes the calibration methodology. Section~\ref{sec:results} presents the empirical results, and Section~\ref{sec:conclusion} concludes.

\section{Bitcoin Inverse Options}
\label{sec:inverse}

\subsection{Payoff structure and convexity}

Let $S_t$ denote the BTC/USD spot price and $K$ a USD strike. A standard (cash-settled) European call pays $C_T^{USD}=\max(S_T-K,0)$. An \emph{inverse} call, settled in Bitcoin, divides this payoff by the terminal spot price:
\begin{equation}
    C_T^{BTC} = \frac{\max(S_T-K,0)}{S_T} = \max\!\left(1-\frac{K}{S_T},0\right),
    \label{eq:inverse_call}
\end{equation}
a dimensionless quantity bounded above by 1 BTC, known in the literature as the \emph{point value} \cite{alexander2023}. The analogous inverse put pays $P_T^{BTC}=\max(K/S_T-1,0)$.

The division by $S_T$ introduces a convexity absent from the standard payoff. For an in-the-money inverse put, $P_T^{BTC}=K/S_T-1$, so
\begin{equation}
    \frac{\partial P_T^{BTC}}{\partial S_T} = -\frac{K}{S_T^2}<0, \qquad
    \frac{\partial^2 P_T^{BTC}}{\partial S_T^2} = \frac{2K}{S_T^3}>0:
\end{equation}
the BTC-denominated liability of a short put position accelerates as the price falls, precisely when the BTC-denominated collateral backing the position is losing dollar value. This ``double risk'' is a defining feature of inverse contracts and motivates dedicated hedging analyses \cite{lucic2024}, which lie outside the scope of the present pricing study.

\subsection{Risk-neutral pricing}
\label{subsec:rn_pricing}

Cryptocurrency derivatives markets conventionally set the risk-free rate $r=0$, reflecting the absence of a money market for Bitcoin \cite{alexander2023,deribit_mark}; we adopt this convention throughout. Under the risk-neutral measure $\mathbb{Q}$, the BTC-denominated price of an inverse call is
\begin{equation}
    C_0^{BTC} = \mathbb{E}^{\mathbb{Q}}\!\left[\frac{\max(S_T-K,0)}{S_T}\right],
    \label{eq:inverse_pricing}
\end{equation}
with $C_0^{USD}=S_0\cdot C_0^{BTC}$. No closed-form solution exists once $S$ follows the rBergomi dynamics of Section~\ref{sec:rbergomi}, so \eqref{eq:inverse_pricing} must be evaluated by Monte Carlo. Because the payoff divides by $S_T$, paths with small terminal value contribute disproportionately to the estimator --- an effect most severe for puts, where $\max(K-S_T,0)/S_T$ is largest exactly when $S_T\to0$. This motivates the variance-reduction techniques of Section~\ref{sec:simulation}. Market-quoted implied volatilities $\sigma_{imp}$ are defined, as on Deribit, by inverting the Black--Scholes formula on the USD-equivalent price, $C_0^{USD}=C_{BS}(S_0,K,T,0,\sigma_{imp})$, which permits direct comparison with model-implied volatilities.

\section{Rough Volatility and the Rough Bergomi Model}
\label{sec:rbergomi}

\subsection{Why rough volatility}

Under the Heston model, $dS_t=\sqrt{v_t}S_t\,dW_t^S$, $dv_t=\kappa(\theta-v_t)dt+\sigma_v\sqrt{v_t}dW_t^v$ with $d\langle W^S,W^v\rangle_t=\rho\,dt$, the at-the-money implied volatility skew converges, as $T\to0$, to the finite limit $\psi_{Heston}(T)\to\rho\sigma_v/(2\sqrt{v_0})$ \cite{gatheral2011,forde2009}. Empirical Bitcoin implied volatility surfaces, by contrast, exhibit a short-maturity skew that steepens without bound as maturity shrinks: the smile-steepness coefficient (the slope of $\sigma_{iv}$ on log-moneyness) falls from approximately 65 for 0--7 day options to roughly 25, 6 and 3 for the 7--30, 30--90 and $>90$ day buckets respectively, while ATM implied volatility rises from $\approx46\%$ to $\approx58\%$ over the same maturities. No finite-$\rho\sigma_v$ Heston configuration reproduces this pattern.

Gatheral, Jaisson and Rosenbaum \cite{gatheral2018} showed that equity realised volatility is consistent with a fractional Brownian motion (fBm) of Hurst parameter $H\approx0.1$, far below the $H=0.5$ of standard Brownian motion. A fBm $W^H$ with $H\in(0,1)$ has covariance $\mathbb{E}[W^H_tW^H_s]=\tfrac12(|t|^{2H}+|s|^{2H}-|t-s|^{2H})$ \cite{mandelbrot1968}; for $H<0.5$ increments are negatively correlated (``rough'' paths), while $H=0.5$ recovers Brownian motion. Multifractal Detrended Fluctuation Analysis applied to Bitcoin realised volatility yields estimates consistently below $H=0.15$ across independent studies (Table~\ref{tab:hurst_estimates}), confirming that Bitcoin's volatility is at least as rough as that of major equity indices.

\begin{table}[H]
\centering
\caption{Hurst exponent estimates for Bitcoin realised volatility (physical measure).}
\label{tab:hurst_estimates}
\begin{tabular}{l c c l}
\toprule
\textbf{Study} & \textbf{Sample period} & \textbf{Estimated $H$} & \textbf{Method} \\
\midrule
Takaishi (2020) \cite{takaishi2020} & 2010--2019 & 0.05--0.12 & MF-DFA (+ structure function) \\
Takaishi (2021) \cite{takaishi2021} & 2011--2020 & 0.08--0.15 & MF-DFA, rolling window \\
Takaishi (2025) \cite{takaishi2025} & 2015--2024 & 0.07--0.14 & MF-DFA, finite-sample corrected \\
\bottomrule
\end{tabular}
\end{table}

\subsection{Model specification}

The rough Bergomi model \cite{bayer2016} specifies, under $\mathbb{Q}$,
\begin{align}
    \frac{dS_t}{S_t} &= \sqrt{v_t}\,dW_t^1, \qquad S_0=s_0, \label{eq:rbergomi_S}\\
    v_t &= \xi_0(t)\,\mathcal{E}\!\left(\eta\int_0^t (t-s)^{H-1/2}\,dW_s^2\right), \qquad v_0=\xi_0(0), \label{eq:rbergomi_v}
\end{align}
where the risk-free drift in \eqref{eq:rbergomi_S} is zero by the convention of Section~\ref{subsec:rn_pricing}, $\mathcal{E}(X_t)=\exp(X_t-\tfrac12\langle X\rangle_t)$ is the stochastic exponential, $d\langle W^1,W^2\rangle_t=\rho\,dt$ with $\rho\in[-1,1]$, $H\in(0,1/2)$ is the Hurst parameter, $\eta>0$ controls the volatility of volatility, and $\xi_0(t)=\mathbb{E}^{\mathbb{Q}}[v_t\mid\mathcal{F}_0]$ is the \emph{forward variance curve} observed at time zero. The kernel $(t-s)^{H-1/2}$ is singular at $s=t$ for $H<1/2$ but square-integrable, producing a Volterra (Riemann--Liouville) process $\widetilde{W}_t=\int_0^t(t-s)^{H-1/2}dW_s^2$ that is Hölder continuous of any order $\gamma<H$ \cite{decreusefond1999}. By the martingale property of the stochastic exponential, $\mathbb{E}^{\mathbb{Q}}[v_t]=\xi_0(t)$ identically: the model is calibrated to the variance term structure \emph{by construction}, leaving only the three shape parameters $(H,\eta,\rho)$ to be estimated from the smile.

Since $\widetilde{W}_t$ is centred Gaussian with $\mathrm{Var}(\widetilde{W}_t)=t^{2H}/(2H)$ (Itô isometry, \cite{nualart2006}),
\begin{equation}
    v_t = \xi_0(t)\exp\!\left(\eta\widetilde{W}_t-\frac{\eta^2 t^{2H}}{4H}\right),
    \label{eq:variance_explicit}
\end{equation}
a log-normal structure analogous to classical stochastic volatility models, but with $\widetilde{W}_t$ rough rather than Brownian. The correlation $\rho<0$ generates the leverage effect: writing $W^2=\rho W^1+\sqrt{1-\rho^2}B$ with $B\perp W^1$, downward moves in $S$ are associated with upward moves in $\widetilde{W}$, steepening the put skew.

\begin{proposition}[ATM skew asymptotics, \cite{bayer2016}]
\label{prop:atm_skew}
Under \eqref{eq:rbergomi_S}--\eqref{eq:rbergomi_v}, the at-the-money skew $\psi(T)=\partial\sigma_{imp}/\partial k|_{k=0}$, $k=\log(K/S_0)$, satisfies
\begin{equation}
    \psi(T) \sim \rho\,\eta\,C_H\,T^{H-1/2} \qquad \text{as } T\to0,
    \label{eq:skew_asymptotics}
\end{equation}
for a constant $C_H>0$ depending only on $H$.
\end{proposition}

For $H<1/2$, $|\psi(T)|\to\infty$ as $T\to0$ --- the power-law explosion observed empirically and unreproducible under Heston. Table~\ref{tab:rbergomi_params} summarises the role of each parameter; typical literature ranges for cryptocurrency markets are $H\in[0.05,0.15]$, $\eta\in[1.5,3.5]$, $\rho\in[-0.9,-0.5]$ \cite{bayer2016,mccrickerd2018}, though our calibration bounds (Section~\ref{subsec:optimisation}) are set wider.

\begin{table}[H]
\centering
\caption{rBergomi parameters and their effect on the implied volatility surface.}
\label{tab:rbergomi_params}
\begin{tabularx}{\textwidth}{c l X}
\toprule
\textbf{Parameter} & \textbf{Name} & \textbf{Effect} \\
\midrule
$H\in(0,0.5)$ & Hurst exponent & Controls the term structure of the skew; lower $H$ $\Rightarrow$ steeper short-maturity skew, decaying faster with $T$. \\
$\eta>0$ & Vol-of-vol & Controls smile curvature/width across all maturities. \\
$\rho\in[-1,0]$ & Spot-vol correlation & Controls skew direction/magnitude; more negative $\rho$ $\Rightarrow$ steeper put skew. \\
\bottomrule
\end{tabularx}
\end{table}

\section{Simulation and Pricing Methods}
\label{sec:simulation}

\subsection{The Hybrid Scheme}

Exact simulation of $\widetilde{W}$ on a grid of $n$ points via Cholesky factorisation of its $n\times n$ covariance matrix costs $O(n^3)$, which becomes prohibitive for $n\gtrsim10^3$. A \emph{coarse-grid Cholesky} scheme mitigates this by factoring the joint covariance of $(\widetilde{W},W^1)$ on a coarse grid of $n_c\ll n$ points and linearly interpolating onto the fine grid, at a cost $O((2n_c)^3+n\,n_c)$ but with an interpolation error.

The \emph{Hybrid Scheme} of Bennedsen et al.\ \cite{bennedsen2017} instead decomposes $\widetilde{W}_{t_i}=R_i+D_i$ into a \emph{recent} component $R_i$, covering the last $\kappa$ steps and simulated exactly via kernel weights $w_l=\frac{1}{H+1/2}\left[(l\Delta)^{H+1/2}-((l-1)\Delta)^{H+1/2}\right]$, and a \emph{distant} component $D_i\approx\sum_{j=0}^{i-\kappa-1}(t_i-t_j)^{H-1/2}\Delta W^2_{j+1}$ approximated by a Riemann sum. This avoids interpolation entirely and operates directly on the fine grid at cost $O(n\kappa)$. The strong approximation error is $O(n^{-H})$ \cite{bennedsen2017}, slower than the classical $O(n^{-1/2})$ rate but adequate in practice for $n=500$--$2000$ \cite{mccrickerd2018}. Algorithm~\ref{alg:hybrid} summarises the full path-generation procedure, including the correlated Brownian increments and the log-Euler update of $S$.

\begin{algorithm}[H]
\caption{Hybrid Scheme path generation for rBergomi}
\label{alg:hybrid}
\begin{algorithmic}
\REQUIRE $(H,\eta,\rho,\xi_0)$; grid $\{t_i\}_{i=0}^n$, step $\Delta$; truncation $\kappa$; paths $N$
\STATE Precompute weights $w_l$, $l=1,\dots,\kappa$, and variance corrections $\alpha_i=\eta^2 t_i^{2H}/(4H)$
\FOR{$m=1$ to $N$}
  \STATE $S_0^{(m)}\gets S_0$, $D_0\gets0$
  \STATE draw i.i.d.\ $\{Z_i^{(1)},Z_i^{(2)}\}_{i=1}^n\sim N(0,1)$
  \FOR{$i=1$ to $n$}
    \STATE $\Delta W_i^1\gets\sqrt{\Delta}Z_i^{(1)}$, $\quad\Delta W_i^2\gets\rho\sqrt{\Delta}Z_i^{(1)}+\sqrt{1-\rho^2}\sqrt{\Delta}Z_i^{(2)}$
    \STATE $R_i\gets\sum_{l=1}^{\min(i,\kappa)} w_l\,Z^2_{i-l+1}$, with $Z^2_j=\rho Z_j^{(1)}+\sqrt{1-\rho^2}Z_j^{(2)}$
    \STATE if $i>\kappa$: $D_i\gets\sum_{j=0}^{i-\kappa-1}(t_i-t_j)^{H-1/2}\Delta W^2_{j+1}$
    \STATE $\widetilde{W}_{t_i}\gets R_i+D_i$, $\quad v_{t_i}\gets\xi_0(t_i)\exp(\eta\widetilde{W}_{t_i}-\alpha_i)$
    \STATE $S_{t_i}^{(m)}\gets S_{t_{i-1}}^{(m)}\exp\left(-\tfrac12 v_{t_{i-1}}\Delta+\sqrt{v_{t_{i-1}}}\Delta W_i^1\right)$
  \ENDFOR
\ENDFOR
\STATE \textbf{return} $\{S_T^{(m)}\}_{m=1}^N$
\end{algorithmic}
\end{algorithm}

\subsection{The Mixed Estimator}
\label{subsec:mixed}

The naive estimator $\hat C_0^{BTC}=\frac1N\sum_m\max(S_T^{(m)}-K,0)/S_T^{(m)}$ has high variance because of the division by $S_T^{(m)}$. Romano and Touzi \cite{romano1997} observed that, conditional on the variance path $\{v_t\}_{t\in[0,T]}$, the log-return is Gaussian: with $I_V=\int_0^Tv_t\,dt$,
\begin{equation}
    \log(S_T/S_0)\mid\{v_t\} \sim \mathcal{N}\!\left(-\tfrac12\rho^2 I_V+\rho\!\int_0^T\!\sqrt{v_t}\,dW_t^2,\ (1-\rho^2)I_V\right).
\end{equation}
For each simulated variance path, the conditional inverse-put price can therefore be computed \emph{analytically} via Black--Scholes applied to the conditional forward $F_m=S_0\exp(-\tfrac12\rho^2I_V^{(m)}+\rho I_{corr}^{(m)})$ and conditional volatility $\sigma_c^{(m)}=\sqrt{(1-\rho^2)I_V^{(m)}}$, eliminating all sampling noise from the independent driver $B_t$. McCrickerd and Pakkanen \cite{mccrickerd2018} report a $5$--$10\times$ variance reduction from this \emph{conditional} estimator alone, and a further $10$--$20\times$ cumulative reduction when it is combined, as a control variate, with a timer-option payoff $Y_m=\mathrm{BS}_{inv}(\max(0,\rho^2(Q_{\max}-I_V^{(m)}));F_m,K)$, $Q_{\max}=\max_mI_V^{(m)}$, yielding the \emph{Mixed Estimator} $Z_m=X_m+\hat\alpha(Y_m-\bar Y)$ with $\hat\alpha=-\mathrm{Cov}(X,Y)/\mathrm{Var}(Y)$ estimated by post-hoc regression.

\subsection{Three computational approaches}
\label{subsec:three_methods}

Combining the two simulation schemes with the two estimators (the Mixed Estimator requires fine-grid Brownian increments, which only the Hybrid Scheme exposes) yields three viable pipelines, summarised in Table~\ref{tab:three_methods}. All three share the same calibration infrastructure (Section~\ref{sec:calibration}) and differ \emph{only} in how variance paths are simulated and how prices are computed from them.

\begin{table}[H]
\centering
\caption{The three computational approaches compared in this paper.}
\label{tab:three_methods}
\begin{tabularx}{\textwidth}{l X X l}
\toprule
\textbf{Method} & \textbf{Scheme} & \textbf{Estimator} & \textbf{Complexity} \\
\midrule
Cholesky+Euler & Coarse-grid Cholesky & Log-Euler MC & $O(n_c^3+n\,n_c)$ \\
Hybrid+Euler & Hybrid Scheme ($\kappa=6$) & Log-Euler MC & $O(n\kappa)$ \\
Hybrid+Mixed & Hybrid Scheme ($\kappa=6$) & Mixed (conditional MC + control variate, antithetic) & $O(n\kappa)$ \\
\bottomrule
\end{tabularx}
\end{table}

\section{Calibration Methodology}
\label{sec:calibration}

\subsection{Data and implied volatility surface extraction}

Trade-level data for all BTC options listed on Deribit between January 2022 and December 2025 ($\approx57{,}000$ instrument files) were obtained via Deribit's public historical API. For each calibration date $t^*$, all trades within $\pm4$ hours of $t^*$ are collected; instruments with traded volume below $0.05$ BTC are discarded, maturities are restricted to $[7,90]$ days (excluding the gamma-dominated, microstructure-noisy sub-week tenors and the illiquid long end), and moneyness $K/S_0$ is restricted to $[0.8,1.2]$. When multiple trades for an instrument fall in the window, implied volatilities are aggregated by volume-weighted average; for each $(K,T)$ pair only the out-of-the-money side (puts for $K<S_0$, calls for $K>S_0$) is retained, as these carry the highest vega-to-price ratio \cite{gatheral2011}. A snapshot is valid if it contains at least 15 distinct $(K,T)$ points across at least 3 maturities and 30 underlying trades.

\subsection{Forward variance curve}

For each maturity $T_k$ present in the surface, the at-the-money implied volatility $\hat\sigma_{ATM}(T_k)$ is identified by a delta-neutral criterion: the option minimising $|d_1(K,T_k)|$ (computed with its own market IV and $r=0$), subject to $|d_1|<0.5$, with the closest-to-spot strike as fallback. The forward variance curve is then
\begin{equation}
    \xi_0(T_k)=\hat\sigma_{ATM}(T_k)^2, \qquad \xi_0(t)=\xi_0(T_k)\ \text{for } T_k\le t<T_{k+1},
    \label{eq:xi0}
\end{equation}
piecewise-constant and extended flatly beyond $[T_1,T_{N_T}]$. This pins down the variance term structure exactly, leaving $(H,\eta,\rho)$ to fit the skew and curvature.

\subsection{Loss function and optimisation pipeline}
\label{subsec:optimisation}

For a candidate $\boldsymbol\theta=(H,\eta,\rho)$, model implied volatilities $\sigma_i^{mod}(\boldsymbol\theta)$ are obtained by (i) simulating $N_{MC}=10{,}000$ variance paths per maturity with $n=100$ time steps under \eqref{eq:rbergomi_v} and the pre-estimated $\xi_0(\cdot)$, using the scheme of Table~\ref{tab:three_methods}; (ii) pricing each inverse put with the corresponding estimator; (iii) inverting via Brent's method ($\sigma\in[0.01,5.0]$, tolerance $10^{-12}$). The calibration objective is the maturity-weighted RMSE (in IV percentage points)
\begin{equation}
    \mathcal{L}(\boldsymbol\theta) = \sqrt{\sum_{i=1}^N w_i\left(\sigma_i^{mod}(\boldsymbol\theta)-\hat\sigma_i\right)^2}\times100, \qquad
    w_i = \frac{1/\sqrt{T_i}}{\sum_{j=1}^N 1/\sqrt{T_j}},
    \label{eq:loss}
\end{equation}
where the $1/\sqrt{T_i}$ weighting follows from the skew asymptotics \eqref{eq:skew_asymptotics}: in the conservative limit $H\to0$, short maturities carry the largest discriminating power for $H$, while the normalisation keeps longer maturities informative for $\eta,\rho$. The unweighted RMSE (the same expression with $w_i=1/N$) is used throughout Section~\ref{sec:results} as the reported, model-independent accuracy metric.

The parameter space $H\in[0.01,0.49]$, $\eta\in[0.3,4.0]$, $\rho\in[-0.99,0.0]$ is searched by Differential Evolution \cite{storn1997} (population 12, max.\ 25 generations, tolerance $0.5$\,pp), followed by Nelder--Mead \cite{nelder1965} refinement. All three computational approaches share identical bounds, optimiser settings and market data; Monte Carlo kernels are JIT-compiled (Numba) with a fixed seed for reproducibility.

\section{Empirical Results}
\label{sec:results}

\subsection{Snapshot design}

Thirty snapshots span May 2022 to March 2025: for each of seven major market events (LUNA/UST collapse, FTX bankruptcy, SVB crisis, spot Bitcoin ETF approval, fourth halving, BTC \$100k milestone, Trump inauguration) the day before, of, and after the event are calibrated (21 snapshots), plus nine baseline dates stratified into low ($\sigma_{ATM}\lesssim45\%$), medium ($45$--$70\%$) and high ($\gtrsim70\%$) implied-volatility regimes (3 dates each).

\subsection{Cross-method comparison}
\label{subsec:cross_method}

Table~\ref{tab:cross_method} reports the aggregate results across all 30 snapshots. Hybrid+Mixed achieves the lowest mean (22.83\,pp) and median (19.60\,pp) RMSE and is simultaneously the fastest (17\,s/snapshot, a $19.9\times$ speed-up over Cholesky+Euler's 5.6 minutes). The advantage is concentrated in event snapshots (22.21\,pp vs.\ 44.90\,pp for Cholesky+Euler); on calm baseline dates the three methods are within $1$\,pp of each other (5.12--14.04\,pp). This counterintuitive joint dominance in speed \emph{and} accuracy arises because the variance reduction of the Mixed Estimator yields a smoother, less noisy loss surface, allowing Differential Evolution to converge in fewer generations.

\begin{table}[H]
\centering
\caption{Aggregate calibration results across 30 snapshots (May 2022--March 2025).}
\label{tab:cross_method}
\footnotesize\setlength{\tabcolsep}{4pt}
\begin{tabular}{l r r r r r}
\toprule
\textbf{Method} & \textbf{Mean RMSE} & \textbf{Median RMSE} & \textbf{Event RMSE} & \textbf{Calm RMSE} & \textbf{Time} \\
\midrule
Cholesky+Euler & 41.76\,pp & 27.68\,pp & 44.90\,pp & 5.12\,pp & 5.6\,min \\
Hybrid+Euler   & 39.43\,pp & 33.67\,pp & 38.94\,pp & 14.04\,pp & 38\,s \\
Hybrid+Mixed   & \textbf{22.83\,pp} & \textbf{19.60\,pp} & \textbf{22.21\,pp} & 5.61\,pp & \textbf{17\,s} \\
\bottomrule
\end{tabular}
\end{table}

\subsection{Event-day stress tests}

The FTX bankruptcy (Table~\ref{tab:ftx}) provides the starkest illustration. On the event day, Cholesky+Euler effectively fails to fit the surface (RMSE $124.65$\,pp), while Hybrid+Mixed achieves $18.84$\,pp; the Trump inauguration shows a similar fourfold gap ($\approx94$\,pp vs.\ $\approx21$\,pp, not tabulated). On the LUNA/UST collapse (Table~\ref{tab:ftx}, lower panel), Hybrid+Mixed remains in the $19$--$33$\,pp range across all three days versus $30$--$73$\,pp for Cholesky+Euler. Across both events the Hybrid+Mixed calibration pins $H$ at the lower bound $0.01$ on five of six days, reflecting extremely rough volatility during acute stress; the one exception (FTX, Day$+1$, $H=0.49$) illustrates that, even within a single event window, the surface can briefly revert to a smoother regime.

\begin{table}[H]
\centering
\caption{Calibration results for the FTX bankruptcy (Nov.\ 2022) and the LUNA/UST collapse (May 2022).}
\label{tab:ftx}
\small
\begin{tabular}{l l r r r r r}
\toprule
\textbf{Day} & \textbf{Method} & $H$ & $\eta$ & $\rho$ & \textbf{RMSE} & \textbf{Time} \\
\midrule
\multicolumn{7}{l}{\emph{FTX bankruptcy --- 8 Nov.\ 2022}}\\
Day$-1$ & Chol.+Euler & 0.072 & 2.882 & $-0.319$ & 12.53\,pp & 2.1\,min \\
        & Hyb.+Mixed  & 0.010 & 1.888 & $-0.337$ & 13.28\,pp & 16\,s \\
Day     & Chol.+Euler & 0.164 & 3.543 & $-0.354$ & 124.65\,pp & 2.0\,min \\
        & Hyb.+Mixed  & 0.010 & 1.922 & $-0.592$ & \textbf{18.84\,pp} & 14\,s \\
Day$+1$ & Chol.+Euler & 0.219 & 3.847 & $-0.808$ & 55.96\,pp & 4.2\,min \\
        & Hyb.+Mixed  & 0.490 & 2.722 & $-0.655$ & 56.26\,pp & 12\,s \\
\addlinespace
\multicolumn{7}{l}{\emph{LUNA/UST collapse --- 9 May 2022}}\\
Day$-1$ & Chol.+Euler & 0.196 & 3.407 & $-0.321$ & 30.35\,pp & 6.1\,min \\
        & Hyb.+Mixed  & 0.010 & 1.059 & $-0.321$ & 18.94\,pp & 15\,s \\
Day     & Chol.+Euler & 0.113 & 3.396 & $-0.110$ & 73.25\,pp & 7.1\,min \\
        & Hyb.+Mixed  & 0.010 & 1.675 & $-0.572$ & 27.23\,pp & 17\,s \\
Day$+1$ & Chol.+Euler & 0.151 & 3.334 & $-0.004$ & 48.06\,pp & 4.6\,min \\
        & Hyb.+Mixed  & 0.010 & 1.782 & $-0.507$ & 32.57\,pp & 19\,s \\
\bottomrule
\end{tabular}
\end{table}

\subsection{Regime dependence}
\label{subsec:regime}

Across the nine baseline dates, mean Hybrid+Mixed RMSE rises from $5.6$\,pp (low IV) to $15.7$\,pp (medium IV) to $51.5$\,pp (high IV). Across all 30 snapshots, the Pearson correlation between $\sigma_{ATM}$ and calibration RMSE is $r=0.89$ (Spearman $r_S=0.91$): calibration error increases approximately linearly with the level of implied volatility, regardless of method. The best single-date fit in the entire sample is the calm-summer-2023 snapshot ($3.94$\,pp, Hybrid+Mixed); the worst is the post-LUNA-recovery date with $\sigma_{ATM}>100\%$ ($88.61$\,pp Hybrid+Mixed, $102.62$\,pp Cholesky+Euler). The performance gap between methods widens with $\sigma_{ATM}$: within $1$\,pp in calm markets, exceeding $40$\,pp in the most stressed regimes.

\begin{figure}[H]
\centering
\begin{minipage}{0.48\textwidth}
    \centering
    \includegraphics[width=\textwidth]{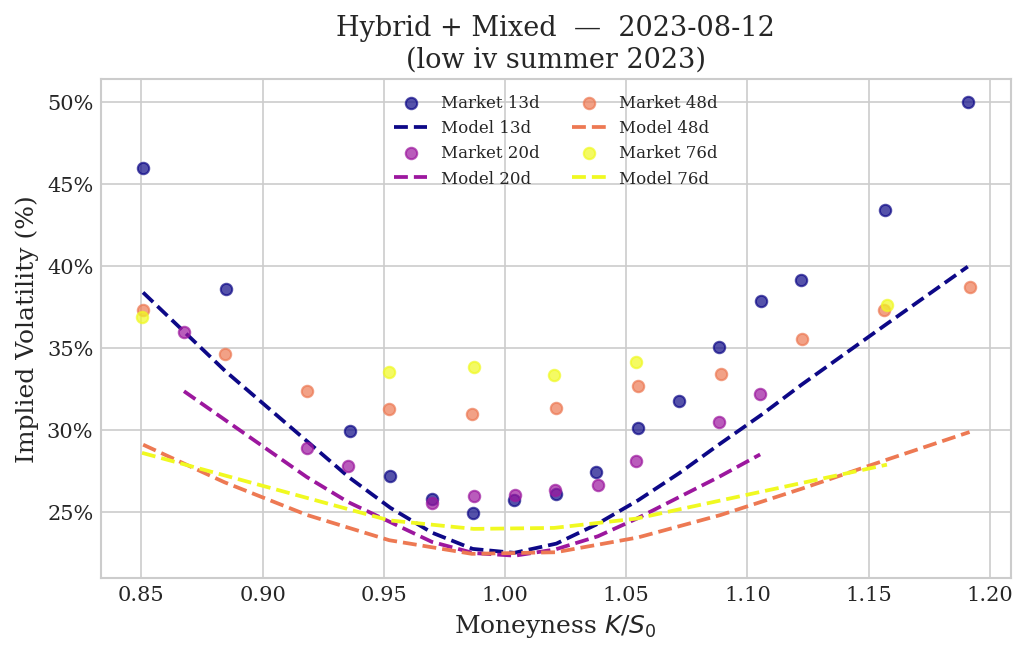}
    \caption*{(a) 12 Aug.\ 2023 (calm), RMSE $=3.94$\,pp}
\end{minipage}
\hfill
\begin{minipage}{0.48\textwidth}
    \centering
    \includegraphics[width=\textwidth]{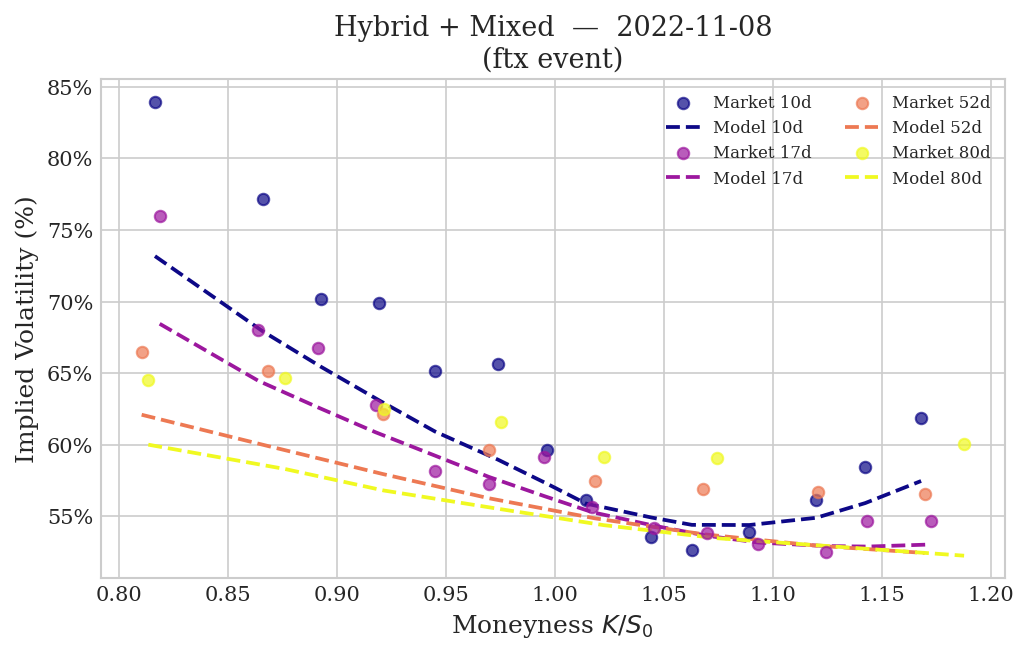}
    \caption*{(b) 8 Nov.\ 2022 (FTX event day), RMSE $=18.84$\,pp}
\end{minipage}
\caption{Hybrid+Mixed calibrated rBergomi smile (dashed) vs.\ market implied volatility (dots) for a calm date and the most extreme event date in the sample. Colours denote distinct maturity slices.}
\label{fig:skew_fits}
\end{figure}

\begin{figure}[H]
\centering
\begin{minipage}{0.48\textwidth}
    \centering
    \includegraphics[width=\textwidth]{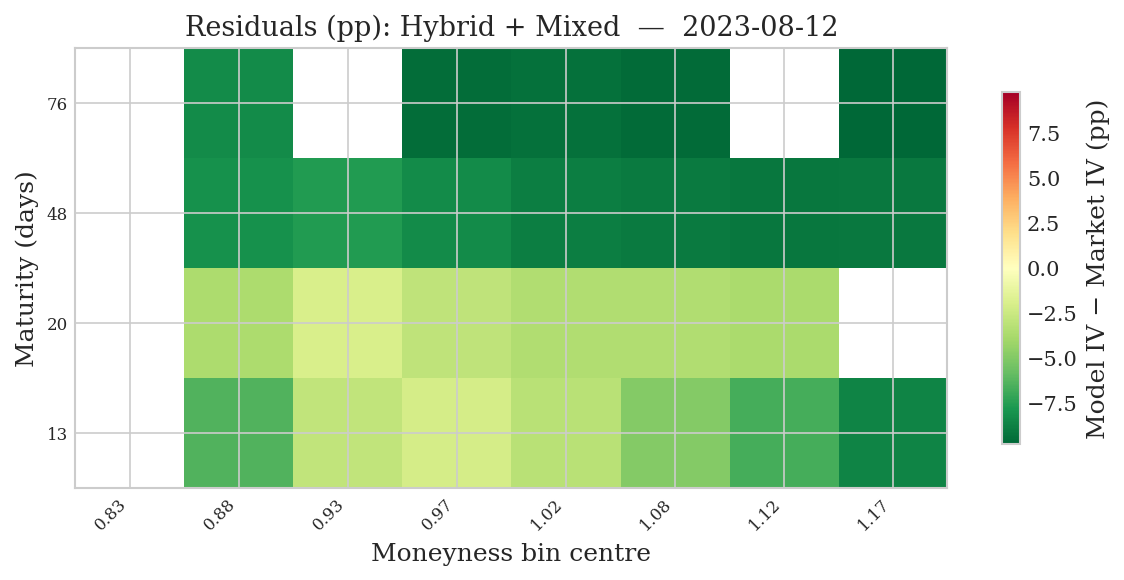}
    \caption*{(a) 12 Aug.\ 2023 (calm)}
\end{minipage}
\hfill
\begin{minipage}{0.48\textwidth}
    \centering
    \includegraphics[width=\textwidth]{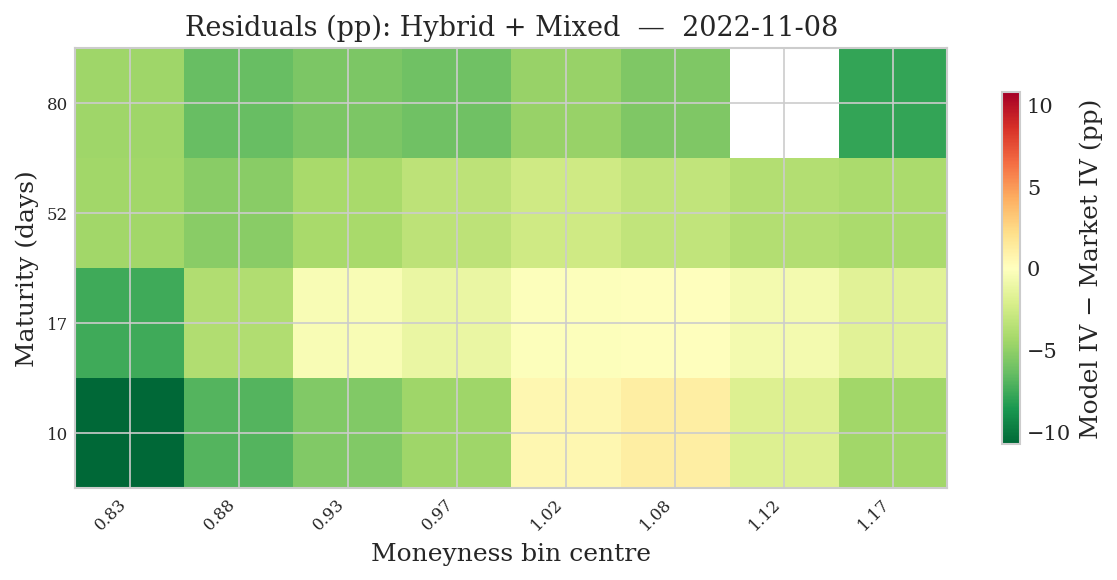}
    \caption*{(b) 8 Nov.\ 2022 (FTX event day)}
\end{minipage}
\caption{Calibration residuals $\varepsilon=\sigma^{mod}-\sigma^{mkt}$ (Hybrid+Mixed) across moneyness and maturity. Green: model below market; red: model above market.}
\label{fig:residuals}
\end{figure}

Figure~\ref{fig:skew_fits} compares the calibrated smile on the best calm-market date and on the FTX event day. Figure~\ref{fig:residuals} shows the corresponding residual heatmaps: in calm conditions residuals are small and structureless, while during the FTX stress a systematic negative bias (model below market) concentrates in the short-maturity, deep out-of-the-money put wing --- the region where the three-parameter rBergomi specification, with $H$ pinned at its lower bound, has exhausted its flexibility to add further roughness.

\subsection{Parameter dynamics}

Across all 30 snapshots, the Hybrid+Mixed calibration yields $H\in[0.01,0.49]$ (mean $0.063$, with the majority of dates at the lower bound $0.01$), $\eta\in[0.30,3.21]$ (mean $\approx1.85$), and $\rho\in[-0.99,0.00]$ (mean $\approx-0.29$, with a substantial fraction of dates at $\rho=0$). The frequent boundary solutions for $H$ and $\rho$ are not pathological: they reflect the optimiser's preference, given a maturity-weighted three-parameter fit, for the roughest admissible dynamics whenever the short-end skew is steep, and for zero correlation whenever the smile is approximately symmetric. The three methods do not converge to the same parameters: Hybrid+Mixed systematically selects lower $H$ and lower $\eta$ (mean $\approx1.85$ vs.\ $\approx3.1$ for Cholesky+Euler) than the other two pipelines, consistent with its cleaner loss surface allowing the optimiser to exploit the short-end skew more aggressively. When $\sigma_{ATM}>70\%$, the calibration systematically favours lower $H$, higher $\eta$, and more negative $\rho$ than in calm regimes ($\sigma_{ATM}<45\%$), where $\eta\approx2.0$--$2.7$ and $\rho\approx0$.

\section{Conclusion}
\label{sec:conclusion}

We have developed a pricing and calibration framework for Bitcoin inverse options under the rough Bergomi model, adapting the BTC-settled payoff $\max(S_T-K,0)/S_T$ to the rBergomi dynamics and benchmarking three computational pipelines --- Cholesky+Euler, Hybrid+Euler, and Hybrid+Mixed --- on thirty Deribit implied volatility surfaces spanning May 2022 to March 2025. The central empirical finding is that the Hybrid+Mixed pipeline, combining the Hybrid Scheme \cite{bennedsen2017} for fractional Brownian motion simulation with the Mixed Estimator \cite{mccrickerd2018} for variance-reduced pricing, is simultaneously the most accurate (mean RMSE 22.83\,pp vs.\ 41.76\,pp for the Cholesky+Euler benchmark) and the fastest (17\,s vs.\ 5.6\,min per snapshot, a 20-fold speed-up), with its advantage concentrated precisely in the high-stress regimes where calibration is hardest. The calibrated Hurst exponent is consistently near the lower bound of the search space, corroborating the rough-volatility hypothesis for Bitcoin under the risk-neutral measure, and calibration error scales approximately linearly with the level of at-the-money implied volatility ($r=0.89$).

The three-parameter rBergomi specification, while parsimonious and robust to estimate, cannot fully reproduce the most distorted surfaces observed during acute crises (RMSE up to $\approx90$\,pp in the most extreme baseline date). Natural extensions include augmenting the model with a jump component or regime-dependent roughness for high-stress periods, evaluating the hedging performance of the calibrated dynamics --- made practically feasible by the computational efficiency of Hybrid+Mixed --- and extending the framework to other cryptocurrency underlyings such as Ethereum.

\section*{Acknowledgements}

The author thanks Prof.\ Giorgia Callegaro (Department of Mathematics ``Tullio Levi-Civita'', University of Padova) for her supervision and guidance throughout the Master's thesis from which this paper is derived.

\bibliographystyle{plain}

\end{document}